\documentclass[manuscript]{acmart}
\usepackage{array}
\usepackage{hhline}
\usepackage{colortbl}
\usepackage[utf8]{inputenc}
\usepackage{color}
\usepackage{multirow}
\usepackage{tabularray}
\AtBeginDocument{%
\providecommand\BibTeX{{%
\normalfont B\kern-0.5em{\scshape i\kern-0.25em b}\kern-0.8em\TeX}}}

\begin{document}
\title[What Makes Digital Support Effective?]{What Makes Digital Support Effective? How Therapeutic Skills Affect Clinical Well-Being}

\author{Wenjie Yang*}
\author{Anna Fang*}
\affiliation{%
 \institution{Carnegie Mellon University}
 \city{Pittsburgh}
 \state{Pennsylvania}
 \country{USA}}

\author{Raj Sanjay Shah}
\affiliation{%
 \institution{Georgia Institute of Technology}
 \country{USA}}

 \author{Yash Mathur}
\affiliation{%
 \institution{Carnegie Mellon University}
 \city{Pittsburgh}
 \state{Pennsylvania}
 \country{USA}}
 
 \author{Diyi Yang}
\affiliation{%
 \institution{Stanford University}
 \country{USA}}

 \author{Haiyi Zhu}
\affiliation{%
 \institution{Carnegie Mellon University}
 \city{Pittsburgh}
 \state{Pennsylvania}
 \country{USA}}

 \author{Robert Kraut}
\affiliation{%
 \institution{Carnegie Mellon University}
 \city{Pittsburgh}
 \state{Pennsylvania}
 \country{USA}}

\renewcommand{\shortauthors}{Fang and Yang et al.}

\begin{abstract}
Online mental health support communities have grown in recent years for providing accessible mental and emotional health support through volunteer counselors. Despite millions of people participating in chat support on these platforms, the clinical effectiveness of these communities on mental health symptoms remains unknown. Furthermore, although volunteers receive some training based on established therapeutic skills studied in face-to-face environments such as active listening and motivational interviewing, it remains understudied how the usage of these skills in this online context affects people's mental health status. In our work, we collaborate with one of the largest online peer support platforms and use both natural language processing and machine learning techniques to measure how one-on-one support chats affect depression and anxiety symptoms. We measure how the techniques and characteristics of support providers, such as using affirmation, empathy, and past experience on the platform, affect support-seekers’ mental health changes. We find that online peer support chats improve both depression and anxiety symptoms \textcolor{black}{with a statistically significant but relatively small effect size. Additionally, support providers' techniques such as emphasizing the autonomy of the client lead to better mental health outcomes.} However, we also found that some behaviors (e.g. persuading) are actually harmful to depression and anxiety outcomes. Our work provides key understanding for mental health care in the online setting and designing training systems for online support providers.

\end{abstract}

\ccsdesc[500]{Human-centered computing~Empirical studies in collaborative and social computing}

\keywords{online communities, mental health, peer support, social computing}

\maketitle

\section{Introduction}
Mental health issues continue to rise globally and under-treatment of serious mental health problems remains a major problem, with more than one in ten people living with a mental health disorder \cite{Dattani2021-he}. Although there is significant evidence supporting the effectiveness of professional treatments, such as therapy, a substantial proportion of people with mental health problems fail to receive any treatment due to reasons such as lacking access to services or having needs unmet by health services \cite{Kim2008-lt, Mojtabai2011-ui}. As a result, peer-to-peer support through online mental health communities (OMHCs) has emerged as an accessible tool for achieving mental and emotional support. OMHCs include sites like 7 Cups and TalkLife, which usually provide free 24/7 peer-to-peer support from volunteers; online support communities are thus able to provide mental and emotional support at scale and on a wide variety of challenges \cite{Dowling2013-ak, Fukkink2011-at}. 

Despite the many benefits available through OMHCs, support providers on online platforms receive relatively little training \cite{Yao2022-ze} in contrast to the extensive training for mental health professionals and even volunteers for crisis intervention programs \cite{Boswell2007-vt, Mufson2006-la, Gould2022-gq, Paukert2004-dx}. OMHCs often require volunteer support providers to complete short training sessions along with optional specialized training; for example, 7Cups.com provides short training sessions based on skills established in face-to-face mental health support such as active listening, empathy, and motivational interviewing (MI) techniques. Another online support platform TalkLife provides small-group and module-led training for peer support techniques including asking questions, showing empathy, and reflection. However, there is little known about whether these therapeutic skills and other “common factors”, which have been researched in non-online contexts \cite{Jani2012-gp, Lundahl2010-bx, Norcross2019-xh}, are actually effective when used by online peer supporters for improving people’s health. This gap on whether therapeutic behaviors are effective in the online context is also especially important due to the fact that they largely inform the design of online support providers’ training. 

\textcolor{black}{As a result, in our work, we collaborate with a large online peer support platform in order to examine the longitudinal and clinical effectiveness of online peer-to-peer support. Using natural language processing and machine learning techniques, we analyze support-seekers' depression and anxiety changes in a dataset spanning over two years containing over 8-million anonymous text-based support chats between volunteer support counselors and online support-seekers.} We explore the effectiveness of support behaviors used by these OMHC volunteers, like asking open-ended questions and providing reflections involved in motivational interviewing, and expressing empathy, and examine how these behaviors are associated with long-term improvement in support-seekers’ depression and anxiety. Our work extends past research, which has largely assessed outcomes of OMHCs on various non-clinical constructs (e.g. mood, retention \cite{Althoff2016-dt,Schueller2017-ms, Wang2023-um}) in cross-sectional studies. Instead, we measure the effectiveness of support chats on changes in clinical outcomes; clinical outcomes in a longitudinal study. Assessing the presence and severity of mental illness symptoms are largely under-studied in the online context despite being the most important outcomes used in important primary care settings, from diagnosis and severity measurement \cite{Kroenke2001-ne, Ruiz2011-kn} to monitoring responses to mental health treatments \cite{Kroenke_Kurt2002-nc, Lowe2004-em}. 

Specifically, our study answers the following research questions:

\textbf{RQ1. Does participating in support chats with volunteers on online peer support platforms improve people’s depression and anxiety?}

\textbf{RQ2. What therapeutic techniques and factors in these conversations affect clinical mental health outcomes?}

\textcolor{black}{Our study found that participating in online peer support chats with volunteer counselors improves symptoms of depression and anxiety with a statistically significant but relatively small effect size.} In particular, having a single online support chat on average improved depression symptom scores by 1.6\% and anxiety symptoms by 0.6\%. Notably, more distressed people who seek help experience even greater improvements. We also found some established therapeutic techniques and behaviors relied on for psychotherapy and in training volunteers in OMHCs are effective while others are harmful. For example, the MI technique of reflecting (capturing and returning something the client has said themselves) is effective for improving depression symptoms, while the MI technique of information-giving actually caused worse symptoms for both depression and anxiety. Overall, our findings provide insight into the efficacy of online support platforms and lend to future work on training for online mental health support.

\section{Related Work}
Below, we review prior work on online mental health communities, therapeutic skills that may lead to better mental health outcomes in both online and offline settings, and success metrics for evaluating OMHCs. 

Table \ref{table:prior-work} situates our work among past literature, which is further reviewed in the sections below. We organize the most relevant prior work and differentiate our study’s use of clinical outcomes, large-scale population data, and evaluation of support providers techniques accordingly.

\subsection{Online Mental Health Communities}
OMHCs include support groups within general-purpose social networks, such as Facebook groups dedicated to specific health issues \cite{Bender2011-fy, Greene2011-gh, Hampton2011-ts, Wellman2001-jv}, and platforms aimed entirely at health support, such as CrisisTextLine \footnote{crisistextline.org}, 7Cups.com \footnote{7cups.com}, and TalkLife \footnote{talklife.com} that connect support seekers with volunteer peer counselors for supportive chats. Peer support through online communities is able to fill gaps in accessing health services for many, helping those whose needs are unmet by traditional resources \cite{Gidugu2015-ds, McBeath2017-ws, Rassau2003-zd} or individuals who lack adequate peer networks in their daily lives to achieve needed support \cite{Kawachi2001-kb, Kessler2005-ky, Solano1986-ey, Turner1983-ri}. Support through OMHCs has been found to have numerous benefits, such as yielding meaningful relationships and increasing trust in getting mental health treatment \cite{Fan2014-kn, Pfeiffer2011-ba, Rogers2015-bb}. Additionally, OMHCs circumvent many key barriers that prevent help-seeking and allow immediate addressing of mental health needs \cite{Demyttenaere2004-bn}. Both general purpose social networks and ones focused exclusively on mental health are important in spreading health information, reducing harmful thoughts, empowering help-seeking for stigmatized populations, reducing suicidal ideation, and fundraising or spreading awareness for health issues \cite{De_Choudhury2014-bj, De_Choudhury2017-xk, Gui2017-tw, Huh2015-nd, Prescott2020-pm}. \textcolor{black}{There is evidence, though, to support that support provided specifically in online mental health platforms may deviate from conventional practices in traditional, offline therapy contexts; for example, there are less distinct boundaries between support-seeker and support-provider, and volunteer counselor behaviors are influenced by training that takes place through platform-provided resources \cite{Yao2022-ze}.}

\subsection{Skills for Counseling and Therapy}
Effective mental health support depends on numerous behaviors and strategies used \cite{Curran2019-ns}. In professional therapy, therapeutic alliance – the extent to which clients and therapists work collaboratively on common goals and connect emotionally – has been the most studied and demonstrably effective element in psychotherapy \cite{Norcross2018-rh, Thomas2005-wd}. Therapist behaviors, including exploration, showing empathy, reflection, accurate interpretation, and active listening, lead to better therapeutic alliance and mental health outcomes \cite{Ackerman2003-ib, Norcross2019-xh, Wampold2023-nx, Norcross2011-jg}. Given that there exist core sets of mental health support skills (“micro-skills”) that transcend specific therapies \cite{Norcross2019-xh}, our work explores how validated counseling techniques (e.g. expressing empathy, motivational interviewing techniques) affect online support-seekers’ clinical outcomes. As noted by Hall and Horvath regarding mental health providers, “micro-skills are the building blocks of effective therapeutic communication … [While they] do not encompass the entire therapeutic skill set, they are useful and accessible in the early stages of counselor training…micro-skills are transtheoretical [and] are the ingredients of the therapeutic process" \cite{Witte_PhD_NCSP2014-rd}. Some of these skills include reflecting on clients’ feelings, asking open-ended questions, and providing empathy \cite{Hill2020-wy, Perez-Rosas2017-lr}. In addition to specific skills, however, generally having a therapeutic space for expression may also be beneficial in itself \cite{Kennedy-Moore2001-ul, Pennebaker2012-dc}.

In online peer counseling, past work has often examined utterance-level skills of support providers \cite{Althoff2016-dt, Chikersal2020-pf, Perez-Rosas2019-no, Zhang2020-kj}. For example, high-quality interactions involve counselors discussing support-seekers' motivations and encouraging them, while lower-quality counseling involves expressing higher persuasion and uncertainty \cite{Perez-Rosas2019-no}. Peer counselors have also been found to have more success when expressing fewer negative emotions and displaying more positive sentiment \cite{Perez-Rosas2019-no, Saha2020-dc}. \textcolor{black}{Our work also contributes to growing work trying to understand how mental health support may differ in the online context compared to the traditional offline context; for example, professional therapists may not be expected to discuss their own issues in traditional therapy sessions, but greater self-disclosure by online peers has been shown to help support-seekers feel more comfortable \cite{Fang2022-gc, Kushner2020-ct, Sharma2018-iv, Vlahovic2014-ws, Wang2012-ny}.}

Below we briefly review frameworks used in our study: (1) empathy, which includes emotional expression, interpretation, and exploration, and (2) motivational interviewing (MI) – a set of validated client-centric counseling techniques used by therapists to help clients make lifestyle changes or work towards recovery \cite{Miller2012-dp, Lundahl2010-bx, Magill2014-qu}. We build on work by Sharma et al. that outlines three forms of empathy – emotional expression, interpretation, and exploration – and uses machine learning to label them \cite{Sharma2020-il}. In addition, we study how online support providers use the different behaviors that are the basis of motivational interviewing, as defined by the Motivational Interviewing Treatment Integrity (MITI) coding manual version 4, which includes various effective, behavior-based techniques for talk therapy including affirmation, open questions, reflections, and informational support \cite{Shah2022-pj, Magill2014-qu}. \textcolor{black}{Some techniques such as affirmation or reflective listening have been found to be consistent with typical therapeutic goals, while other therapist behaviors like confrontation are unlikely to lead to positive mental health changes in clients \cite{RN9261, RN9262}.}

\subsection{Measuring Effectiveness of Online Support Communities}
Over the past few decades, there has been ample research trying to understand the impact and effectiveness of support in online communities. \textcolor{black}{There is a growing perspective over the past few years in HCI research for including holistic models of well-being, such as the biopsychosocial model, which can take into account biological, psychological, and social factors that contribute to a person in order to identify better support methods and/or understand a person's experience better \cite{marcu2023attachment, wade2017biopsychosocial}. While our work contributes a medical perspective of depression and anxiety, we also fully acknowledge that quantitative measurements of mental health symptoms is simply one of many ways for considering one's health, in addition to other holistic factors.}

In terms of evaluating people's well-being from a quantitative perspective, previous work has used a variety of proxies. For example, past work has used client feedback immediately following a counseling session \cite{Fang2022-gc, Wang2023-um, Zhang2019-fi} and their continued engagement on the platform \cite{Sharma2020-il} as indications of high-quality peer support. Wang et al. found that some linguistic predictors are positively related to people giving higher evaluation scores to peer counselors but negatively related to their retention on the platform. Unfortunately, key limitations exist with using these outcome constructs as they have been shown to be unclear in value and meaning in the online mental health context \cite{Cunha2019-mx, Wang2023-um} and likely measure fundamentally different aspects of well-being \cite{Wang2023-um}. Retention to the platform, for example, has been a notable focus of past research to measure the success of online health communities \cite{Chen2021-wv, Wang2012-ny, Yang2017-dl}; however, retention can be complex in interpretation in this context as Massimi et al. found that people may opt not to return to online mental health platforms due to their positive improvement offline \cite{Massimi2014-gk}. Additionally, many of these past measures have been limited to the immediate effects of counseling sessions rather than observing longer-term improvements. Other evaluation metrics have relied on human assessment, such as defining high-quality counseling sessions based on literature and coding these sessions with these predefined criteria \cite{Huang_undated-xb, Perez-Rosas2019-no}; for instance, because reflective listening is a valuable technique in conventional offline therapy, researchers have measured the extent to which reflective listening is occurring online as a metric for the quality of online counseling sessions \cite{Miller2012-dp}. Despite there being relatively little prior work using clinical outcomes to measure improvement from digital mental health support \cite{Chikersal2020-pf, Doherty2012-da}, the gold standard for clinical research in mental health is the use of self-report questionnaires measuring mental health symptoms (e.g. PHQ-9 for depression, GAD-7 for anxiety) or mental health evaluations by trained professionals. While PHQ-9 and GAD-7 are widely adopted and verified assessments in mental health research and practice, low data availability and response rate often make it impossible to study at scale. Our work extends the past literature as shown in Table \ref{table:prior-work}, and analyzes clinical outcomes from chat-based volunteer support in a large-scale online peer support community using short versions of the PHQ and GAD questionnaires for depression and anxiety symptom assessment, respectively.

\begin{table}
\centering
\begin{tabular}{|>{\hspace{0pt}}m{0.105\linewidth}|>{\hspace{0pt}}m{0.11\linewidth}|>{\hspace{0pt}}m{0.13\linewidth}|>{\hspace{0pt}}m{0.08\linewidth}|>{\hspace{0pt}}m{0.14\linewidth}|>{\hspace{0pt}}m{0.1\linewidth}|>{\hspace{0pt}}m{0.11\linewidth}|} 
\cline{2-7}
\multicolumn{1}{>{\hspace{0pt}}m{0.09\linewidth}|}{} & \textbf{Large-scale sample} & \textbf{Sample size} & \textbf{1-on-1}\par{}\textbf{support chats} & \textbf{Volunteer/non-professional support providers} & \textbf{Clinical mental health outcomes} & \textbf{Based on established therapeutic skills}\\ 
\hhline{=======|}
\rowcolor[rgb]{0.886,0.875,0.875} \textbf{Our Work} & $\checkmark$ & 74K people \par{}(depression)\par{}42K people \par{}(anxiety) & $\checkmark$& $\checkmark$ & $\checkmark$ & $\checkmark$\\ 
\hline
Althoff \par{}et al. \cite{Althoff2016-dt} & $\checkmark$ & 15K \par{}conversations& $\checkmark$& & &\\ 
\hline
Chikersal \par{}et al. \cite{Chikersal2020-pf} & $\checkmark$ & 54K \par{}people&& $\checkmark$ & $\checkmark$ &\\ 
\hline
Sharma \par{}et al. \cite{Sharma2020-il}& $\checkmark$ & 34 million posts by\par{}1.3 million people && & &\\ 
\hline
Doherty \par{}et al. \cite{Doherty2012-da} & & 18 people&& $\checkmark$ & $\checkmark$ &\\ 
\hline
Saha \par{}et al. \cite{Saha2020-dc} & $\checkmark$ & 300K posts by\par{}39k people&& $\checkmark$ & &\\ 
\hline
Zhang \par{}et al. \cite{Zhang2019-fi}& $\checkmark$ & 1 million \par{}conversations& $\checkmark$& $\checkmark$ & &\\ 
\hline
Wang \par{}et al. \cite{Wang2023-um}~& $\checkmark$ & 1.7 million \par{}conversations & $\checkmark$& $\checkmark$ & &\\ 
\hline
Pérez-Rosas\par{}et al. \cite{Perez-Rosas2019-no} & & 259\par{} conversations&& & & $\checkmark$\\ 
\hline
Chen \par{}and Xu \cite{Chen2021-wv}& $\checkmark$ & \textgreater{}60K posts by\par{}46K people && $\checkmark$ & &\\
\hline
\end{tabular}
\caption{Relevant prior work classified across dimensions including sample size, platform specifications, mental health outcomes, and analysis predictors.}
\label{table:prior-work}
\end{table}

\section{Data}
Our study’s research site is one of the largest online peer support platforms, which we will refer to anonymously as the "online support platform". The online support platform is an online psychological support service where people (support-seekers) can participate in anonymous text-based chats with volunteer support providers on a variety of mental health problems. We will use the terms “support-seekers” and “support providers” for the remainder of this paper. Our dataset spans January 2020 to May 2022 and contains over 8 million chats, involving over 1.5 million support-seekers, and 288 thousand support providers. A typical (median) chat lasts for 23 minutes and, over the span of our two-year dataset, support-seekers chatted with 12 distinct providers on average. \textcolor{black}{Our data contains all chat messages in a chat session, including both support-seeker and support-provider messages.}\\

\noindent \textbf{Privacy, Ethics, and Disclosure.} \textcolor{black}{This paper used anonymous behavioral log data obtained through a collaboration with an online support platform to conduct our analysis. All data was anonymized throughout the entire research process including analysis and no personally identifiable information was used in this study. No authors on this paper are affiliated with the support platform nor did this research receive any funding from the platform. Our study does not introduce interventions or alter people's experiences on the platform in any way given the potential dangers and ethical concerns in doing so, and we have chosen instead to only use people's prior log data to understand how they currently experience the platform. We also find it worth noting ethical concerns in a popular thread within online mental health research generally centering around predicting mental health status from online community behavior \cite{chancellor2020methods}. For example, prior work has often used vague definitions of mental health (such as generalized terms rather than medical definitions of "anxiety" or "depression") and lack theoretical or clinically rigorous grounding in measuring mental health disorders \cite{chancellor2020methods}. Although mental health can be measured through several different manners, our study hopes to address these rightful concerns through referencing only the clinical definitions of depression and anxiety (as opposed to colloquial or general reference) as well as using the established assessments for these conditions in medical and psychiatry contexts (i.e. PHQ and GAD assessments). Generally, we hope our study contributes to a growing perspective in the field for understanding people's support-seeking and support-providing behaviors, and based on people's self-reported data rather than predicting or labeling individual's statuses or diagnoses. We also take note from \cite{chancellor2019human} about dehumanizing terminology used in research studies about mental health. As a result, in this paper we avoid using the terms "users", "subjects", "participants", "observations", or "accounts" when referring to the people who come to our study's online platform for mental health support; instead, we have chosen the term "support-seeker" (and, at times, "people with depression/anxiety" and "individuals"). Additionally, our team does not name nor consider these online users as "patients" given we are not studying any professional help and characterize the relationship between support-seekers and support-providers on our study's platform as closer to a peer relationship.}

\subsection{Counselor Training}
Support providers are required to complete a roughly one-hour, psychology-based training that is based on active listening and MI skills. Support-provider training includes learning skills such as asking guiding questions, reflecting concerns back to support-seekers, and showing empathy. After this initial training, support providers must pass an exam on the site to begin chatting with support-seekers. Support providers can also receive awards on their profiles from completing additional training modules such as specialized courses for specific conditions (e.g. ADHD, Depression) as well as advanced general skill courses (e.g. “Active Listening”, “Managing Emotions”).

\subsection{Clinical Assessments}
The online support platform encourages support-seekers to take free clinical mental health exams. Support-seekers can complete PHQ-9 and PHQ-2 to measure their depression and GAD-7 and GAD-2 to measure anxiety. These assessment instruments are routinely used in both the research and medical setting for diagnostic and severity measurement \cite{Wang2018-ah, Ruiz2011-kn, Arroll2010-ox, Behar2009-zs, Lowe2006-zl}. PHQ-2 includes the first two questions of PHQ-9, which is the 9-item questionnaire measuring depression presence and severity while, GAD-2 includes the first two questions of GAD-7, the 7-item questionnaire measuring anxiety severity. Support-seekers can take the PHQ and GAD exams every two weeks to measure any changes in their depression and anxiety symptoms. \textcolor{black}{Our study's analysis presented later in this study will use support-seekers' repeated and longitudinal questionnaire results to measure their changes in depression and anxiety over time.}

Although the 2-item versions for PHQ and GAD are not as accurate as the longer questionnaires, PHQ-2 and GAD-2 have been shown in prior work to maintain relatively high accuracy including both considerable sensitivity and specificity for measuring depression and anxiety, respectively \cite{Arroll2010-ox, Lowe2005-kl, Plummer2016-bd}. Sensitivity and specificity for PHQ-2 are 86\% and 78\%, respectively \cite{Arroll2010-ox}; GAD-2 has a sensitivity of 76\% and specificity of 81\% \cite{Plummer2016-bd}. A complete list of items in PHQ-2 and GAD-2 can be found in the appendix. A \textit{higher} score on PHQ and GAD indicate greater severity of symptoms; a lower score indicates better well-being than a higher score. 

\begin{figure}[h!]
\includegraphics[scale=0.25]{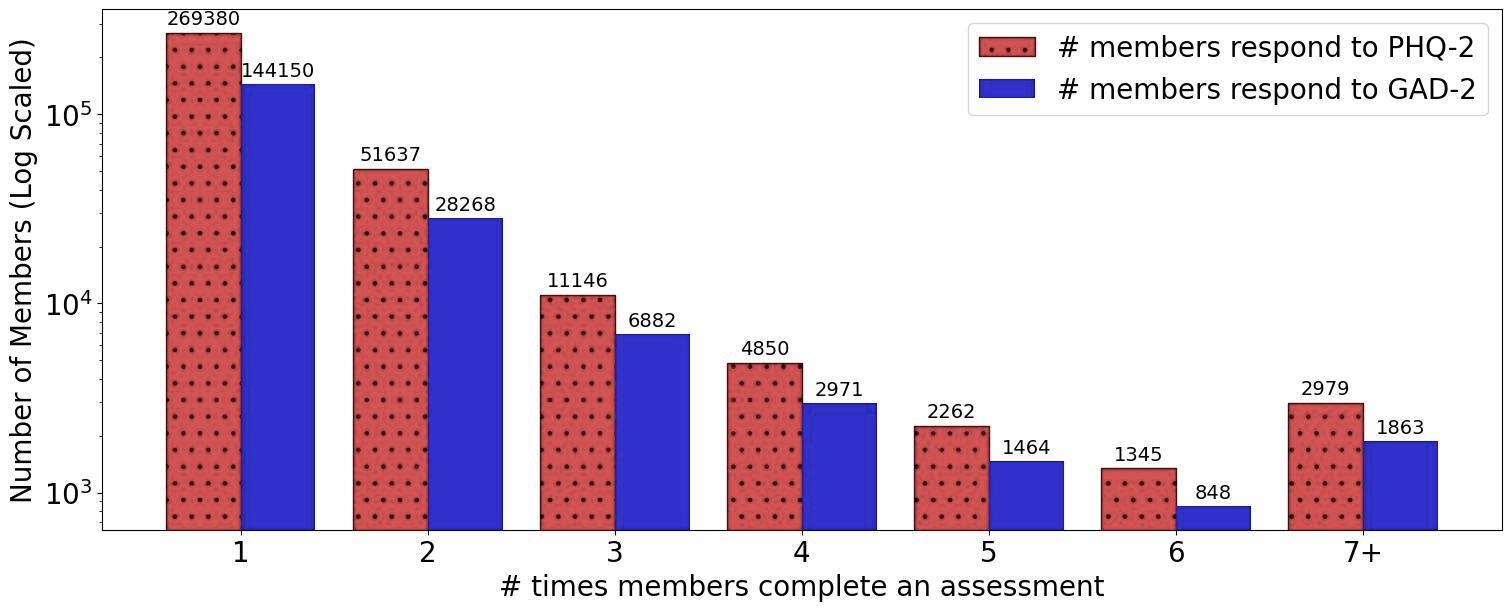}
\caption{Distribution of the number of times support-seekers completed the PHQ-2 and GAD-2 clinical assessments}
\label{figure:assessment-distribution}
\end{figure}

The platform collected 343,599 support-seekers’ responses for PHQ-2 and 186,446 for GAD-2 between January 2020 to May 2022. Because we are interested in changes in mental health, we focus on the 74,219 support-seekers who completed PHQ-2 and 42,296 who completed GAD-2 at least twice. The detailed distribution of the number of responses per support seeker is shown in Figure \ref{figure:assessment-distribution}.

\section{RQ1. Does participating in peer-support chats improve support-seekers’ depression and anxiety?}
We first present our methods and results to answer RQ1: Does participating in peer support chats with lightly trained volunteers on online peer support platforms improve users’ depression and anxiety?

Because PHQ-2 and GAD-2 assessments were administered to all support-seekers independently of whether they had engaged in peer support chats with support providers, it is possible to differentiate between support-seekers who engaged in peer support chats versus those who did not. To answer RQ1, we will compare changes in PHQ-2 and GAD-2 assessments between these two groups through propensity score matching (Section 4.1.1) and conduct analysis on the difference in their mental health assessments using regression analysis (Section 4.1.2).

\subsection{Methods}
\textcolor{black}{Due to ethical concerns, conducting randomized controlled trials or other interventional studies to investigate our research questions was not pursued.} Instead, we adopted a pre-post observational study design to infer the treatment effects of engaging in one or more chats on support-seekers’ mental health status. Specifically, we adopted an “observation-level" random-effects regression with propensity score matching to predict a support-seeker’s post-assessment mental health scores from their pre-assessment score, depending on whether they chatted with a support-provider during the observation period and other independent variables. 

In order to monitor mental health changes, we first selected an “observation period” as our unit of analysis throughout this study. This period is defined by a series of two consecutive PHQ-2 or GAD-2 assessments completed by a support seeker. We refer to the first questionnaire in this sequence as the “pre-assessment”, and the following one as the “post-assessment”. These observation periods enable repeated observations for the same individuals at different time points; thus, this panel data allows us to use multilevel models like random effects to account for potential omitted-variable bias.

\subsubsection{Mitigating Selection Bias via Propensity Score Matching}
To account for individual differences between people that may bias whether support-seekers engage in support chats (e.g. age, gender), we employed Propensity Score Matching (PSM), a quasi-experimental method that emulates the conditions of Randomized Controlled Trials (RCTs) prior to regression analysis. PSM is widely recognized for its robust performance in reducing the effects of confounding in observational studies \cite{austin2011introduction} and has been successfully applied in OMHC scenarios similar to our own \cite{Chen2021-wv}.

In our study, each observation is categorized into the treatment group if the support-seeker engaged in at least one chat during the observation period. \textcolor{black}{Although another analysis method would include using the total number of chats during the observation period rather than splitting support-seekers into just two groups ("had a chat" versus "did not have a chat"), data limitations limit us from splitting "treated" support-seekers into many different (small) groups by their sum number of chats and matching based on this number. However, this method is worth exploring in future work.} The aim of PSM is to match each observation in the treatment group with an observation in the control group (i.e. support-seekers who did not participate in chats) that is highly similar on confounding characteristics such as demographics and mental health status. To achieve this, PSM first estimates a propensity score for every observation – the probability of being in the treatment group conditional on certain covariates – and then pairs observations from the treatment and control groups who have similar propensity scores. This process results in treated and control pairs with similar distribution in terms of the covariates \cite{austin2011introduction} but differing on whether they received “treatment” or not (see Section 2.2). Observations whose propensity scores differ by at most a pre-specified amount (the caliper width) are excluded. 
\\

\noindent\textbf{Step 1. Propensity Score Estimation. }

We adopted a logistic regression model to estimate propensity scores, which predict the probability a support-seeker engages in chats in a given observation period based on the below covariates. 

Although demographic factors of help-seekers such as age, gender, and ethnicity are frequently considered in past studies \cite{dowling2014impacts, fukkink2009children, giorgio2013using, haner2016live, levitz2018influence, rodda2014characteristics}, the OMHC which is our research site collects limited background information about their users. \begin{itemize}
\item \textbf{Support-seekers’ age}: The mean age of support-seekers is 24.52 (±10.61), reflecting the predominantly young user base of this peer support platform. Age is the only demographic variable we can reliably measure because it is a required field seekers must provide when signing up for the online support platform. 
\end{itemize} 

Another key factor is help-seekers’ prior experience on the platform \cite{fukkink2009children}. We operationalize this through three metrics:

\begin{itemize}
\item \textbf{Experience in chats}: the total number of chats support-seekers have participated in from the time of their registration on the platform until the start of the observation period. 
\item \textbf{Tenure}: the number of days since a support-seeker registered on the platform
\item \textbf{Active days}: the total number of days a support-seeker has engaged in any activity on the platform.
\end{itemize}

Lastly, we consider:
\begin{itemize}
\item \textbf{Support-seekers’ initial mental health status}: measured by the support-seeker’s PHQ-2 or GAD-2pre-assessment score.
\end{itemize}

\noindent\textbf{Step 2. Matching and Balance Estimation.}

\begin{table}
\centering
\caption{Covariate balance across treatment and control groups before and after caliper 1:1 matching without replacement on the logit of propensity score.}
\label{table:psm-balance}
\arrayrulecolor{black}
\begin{tabular}{llrrrr|rrrr}
\rowcolor[rgb]{0.996,1,1}                                                                                                                                                            &         & \multicolumn{4}{c|}{PHQ-2}                                                                                                                                                          & \multicolumn{4}{c}{GAD-2}                                                                                                                                                            \\ 
\hhline{>{\arrayrulecolor[rgb]{0.996,1,1}}-->{\arrayrulecolor{black}}--------}
\rowcolor[rgb]{0.996,1,1}                                                                                                                                                            &         & \textbf{Treat.} & \textbf{Control} & \textbf{\%bias} & \begin{tabular}[c]{@{}>{\cellcolor[rgb]{0.996,1,1}}r@{}}\textbf{\%reduct }\\\textbf{ \textbar{}bias\textbar{}}\end{tabular} & \textbf{Treat.} & \textbf{Control} & \textbf{\%bias} & \begin{tabular}[c]{@{}>{\cellcolor[rgb]{0.996,1,1}}r@{}}\textbf{\%reduct }\\\textbf{ \textbar{}bias\textbar{}}\end{tabular}  \\ 
\hhline{>{\arrayrulecolor[rgb]{0.996,1,1}\doublerulesepcolor[rgb]{0.996,1,1}}==>{\arrayrulecolor{black}\doublerulesepcolor{white}}========}
\rowcolor[rgb]{0.996,1,1} \textbf{Covariate}                                                                                                                                         &         &                  &                  &                 & \multicolumn{1}{l|}{}                                                                                                       &                  &                  &                 &                                                                                                                              \\ 
\hhline{=>{\arrayrulecolor[rgb]{0.996,1,1}\doublerulesepcolor[rgb]{0.996,1,1}}=========}\doublerulesepcolor{white}
\rowcolor[rgb]{0.996,1,1} {\cellcolor[rgb]{0.996,1,1}}                                                                                                                               & Raw     & 23.4             & 25.3             & -16.5           & {\cellcolor[rgb]{0.996,1,1}}                                                                                                & 23.4             & 25.3             & -16             & {\cellcolor[rgb]{0.996,1,1}}                                                                                                 \\
\rowcolor[rgb]{0.996,1,1} \multirow{-2}{*}{{\cellcolor[rgb]{0.996,1,1}}\begin{tabular}[c]{@{}>{\cellcolor[rgb]{0.996,1,1}}l@{}}Support-seeker\\ age\end{tabular}}                    & Matched & 23.5             & 23.4             & 2.0             & \multirow{-2}{*}{{\cellcolor[rgb]{0.996,1,1}}88.1}                                                                          & 23.6             & 23.5             & 1.5             & \multirow{-2}{*}{{\cellcolor[rgb]{0.996,1,1}}90.6}                                                                           \\ 
\arrayrulecolor{black}\hline
\rowcolor[rgb]{0.996,1,1} {\cellcolor[rgb]{0.996,1,1}}                                                                                                                               & Raw     & 3.5              & 3.5              & 0.0             & {\cellcolor[rgb]{0.996,1,1}}                                                                                                & 3.8              & 3.7              & 3.4             & {\cellcolor[rgb]{0.996,1,1}}                                                                                                 \\
\rowcolor[rgb]{0.996,1,1} \multirow{-2}{*}{{\cellcolor[rgb]{0.996,1,1}}\begin{tabular}[c]{@{}>{\cellcolor[rgb]{0.996,1,1}}l@{}}Pre-assessment\\ mental health \\ score\end{tabular}} & Matched & 3.6              & 3.6              & 1.2             & \multirow{-2}{*}{{\cellcolor[rgb]{0.996,1,1}}-2452.1}                                                                       & 3.8              & 3.8              & -0.2            & \multirow{-2}{*}{{\cellcolor[rgb]{0.996,1,1}}95.0}                                                                           \\ 
\hline
\rowcolor[rgb]{0.996,1,1} {\cellcolor[rgb]{0.996,1,1}}                                                                                                                               & Raw     & 13.3             & 2.6              & 67.7            & {\cellcolor[rgb]{0.996,1,1}}                                                                                                & 13.9             & 2.7              & 72.6            & {\cellcolor[rgb]{0.996,1,1}}                                                                                                 \\
\rowcolor[rgb]{0.996,1,1} \multirow{-2}{*}{{\cellcolor[rgb]{0.996,1,1}}\begin{tabular}[c]{@{}>{\cellcolor[rgb]{0.996,1,1}}l@{}}Support-seeker~\\experience in chats\end{tabular}}    & Matched & 4.4              & 3.5              & -2.3            & \multirow{-2}{*}{{\cellcolor[rgb]{0.996,1,1}}96.5}                                                                          & 3.5              & 4.3            & -2.1            & \multirow{-2}{*}{{\cellcolor[rgb]{0.996,1,1}}97.1}                                                                           \\ 
\hline
\rowcolor[rgb]{0.996,1,1} {\cellcolor[rgb]{0.996,1,1}}                                                                                                                               & Raw     & 326.0            & 317.9            & 12.5            & {\cellcolor[rgb]{0.996,1,1}}                                                                                                & 373.0            & 364.5            & 11.8            & {\cellcolor[rgb]{0.996,1,1}}                                                                                                 \\
\rowcolor[rgb]{0.996,1,1} \multirow{-2}{*}{{\cellcolor[rgb]{0.996,1,1}}\begin{tabular}[c]{@{}>{\cellcolor[rgb]{0.996,1,1}}l@{}}Support-seeker\\ tenure\end{tabular}}                 & Matched & 285.9            & 277.2            & 0.3             & \multirow{-2}{*}{{\cellcolor[rgb]{0.996,1,1}}97.6}                                                                          & 320.3            & 310.4            & 1.7             & \multirow{-2}{*}{{\cellcolor[rgb]{0.996,1,1}}85.4}                                                                           \\ 
\hline
\rowcolor[rgb]{0.996,1,1} {\cellcolor[rgb]{0.996,1,1}}                                                                                                                               & Raw     & 19.9             & 22.6             & 18.4            & {\cellcolor[rgb]{0.996,1,1}}                                                                                                & 21.2             & 28.3             & 12              & {\cellcolor[rgb]{0.996,1,1}}                                                                                                 \\
\rowcolor[rgb]{0.996,1,1} \multirow{-2}{*}{{\cellcolor[rgb]{0.996,1,1}}\begin{tabular}[c]{@{}>{\cellcolor[rgb]{0.996,1,1}}l@{}}Support-seeker's\\ active days\end{tabular}}          & Matched & 10.2             & 13.3             & -0.6            & \multirow{-2}{*}{{\cellcolor[rgb]{0.996,1,1}}96.8}                                                                          & 11.3             & 14.3             & 0.8             & \multirow{-2}{*}{{\cellcolor[rgb]{0.996,1,1}}93.1}                                                                          
\end{tabular}
\end{table}

\begin{figure}[!h]
    \centering
    \includegraphics[width=\linewidth]{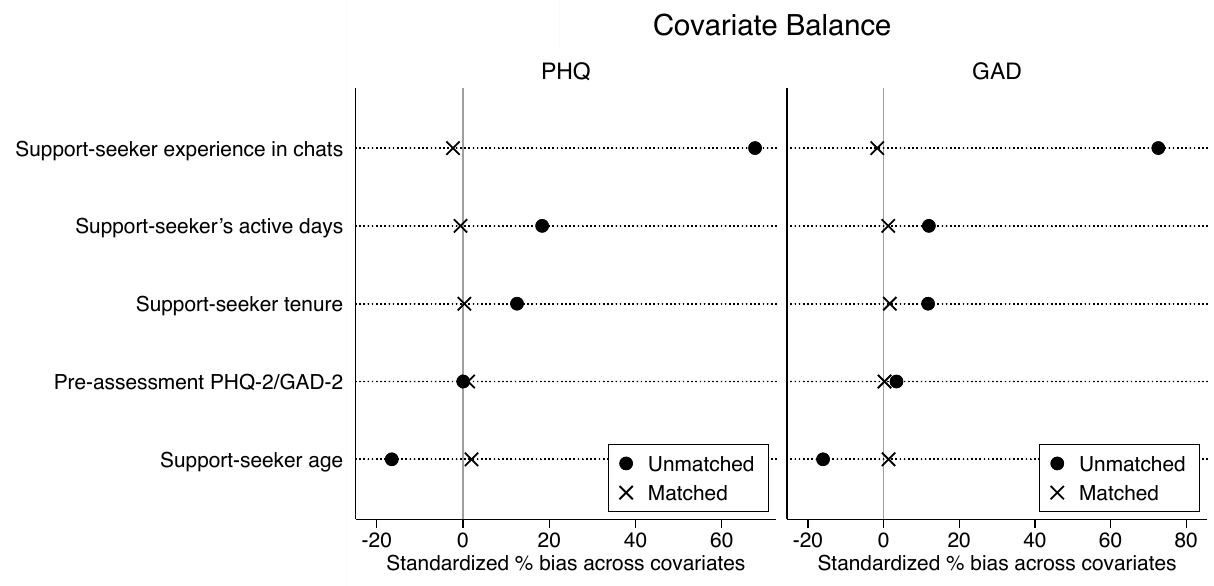}
  \caption{Love plot showing covariate balance for PHQ (left) and GAD (right) before ("Unmatched") and after ("Matched") matching process.}
    \label{figure:loveplot}
\end{figure}

We employed a one-on-one non-replacement matching strategy, based on the logit of the estimated propensity score. We paired each observation in the treatment group with a counterpart in the control group that has the closest logit within the defined caliper. The caliper was set at 0.2 of the standard deviation of the propensity score’s logit, following Austin’s recommendations \cite{austin2011optimal} -- specifically, 0.205 for PHQ outcomes and 0.226 for GAD outcomes. 

Following matching, we identified 43,249 matched pairs for PHQ with 12,854 people in the treatment group unable to find a match and thus removed from this analysis. For GAD, the matched pairs are 26,224; 9,747 people in the treatment group were removed from this analysis. Table \ref{table:psm-balance} details the balance evaluation after matching. Most of the variables observe an 80-90\% reduction in bias following matching, and all the covariates exhibit a bias of less than five, the threshold recommended by Austin \cite{austin2011optimal}. 

\subsubsection{Regression Analysis}
After the matching process, we conducted a random-effects, multi-level regression with observations nested within support-seeker. Because approximately half of the sample had only a single observation, fixed effects regressions would not be appropriate. This regression predicts support-seekers post-assessment mental health (i.e., PHQ and GAD scores) from their pre-assessment mental health, dependent on whether they had at least one support chat during the observation period (treatment group) or not (control groups). Including the support-seeker's \textbf{pre-assessment mental health scores} (i.e., PHQ-2 or GAD-2) in the regression models effectively examines how the independent variables predict changes in mental health status \cite{cohen2013statistical}. To test whether participating in a support-chat has stronger effects for those with worse mental health, we include the interaction between participating in a chat with pre-assessment mental health. We also include as a control variable the number of support-chats the seeker had before the observation.

\subsection{Results}

\begin{table}[!ht]
    \centering
    \begin{tabular}{|l|cc|cc|}
   \toprule
        Predicting Member's Post-Assessment Mental Health & ~ & ~ & ~ & ~ \\ 
        \hline
        ~ & \multicolumn{2}{c|}{PHQ-2}& \multicolumn{2}{c|}{GAD-2} \\ 
        ~ & Coefficient & Robust SE & Coefficient & Robust SE \\ 
        \hline
        Support-seeker experience in chats (log) & -0.052*** & 0.006 & -0.065*** & 0.008 \\


        Pre-assessment mental health score (MHS) & 0.668*** & 0.004 & 0.683*** & 0.005 \\
        Treatment: Participated in $\ge 1$ chats during observation period & -0.126*** & 0.01 & -0.036** & 0.013 \\ 
       Treatment $\times$ MHS & -0.155*** & 0.006 & -0.145*** & 0.007 \\ 
        Constant & 3.620*** & 0.007 & 3.739*** & 0.01 \\ 
        \midrule
        N\_respondents & 53915 & ~ & 32222 & ~ \\ 
        N\_obs & 86498 & ~ & 52448 & ~ \\ 
        Rsq\_overall & ~ & ~ & ~ & ~ \\ 
        $* p< 0.05;$ $** p<0.01;$  $*** p<0.001$ & & & &\\ 
    \bottomrule
        \end{tabular}
\caption{Random-effects regression predicting support-seekers’ post-assessment PHQ-2 or GAD-2 scores, comparing matched seekers who had or did not have a support chat during the observation period, controlling for their pre-assessment mental health and the number of chats they had prior to the observation period and the pre-assessment mental health X number of prior chat interactions.  
\label{RQ1-results}
}\end{table}



\begin{figure}[!h]
    \centering
    \includegraphics[scale=.2]{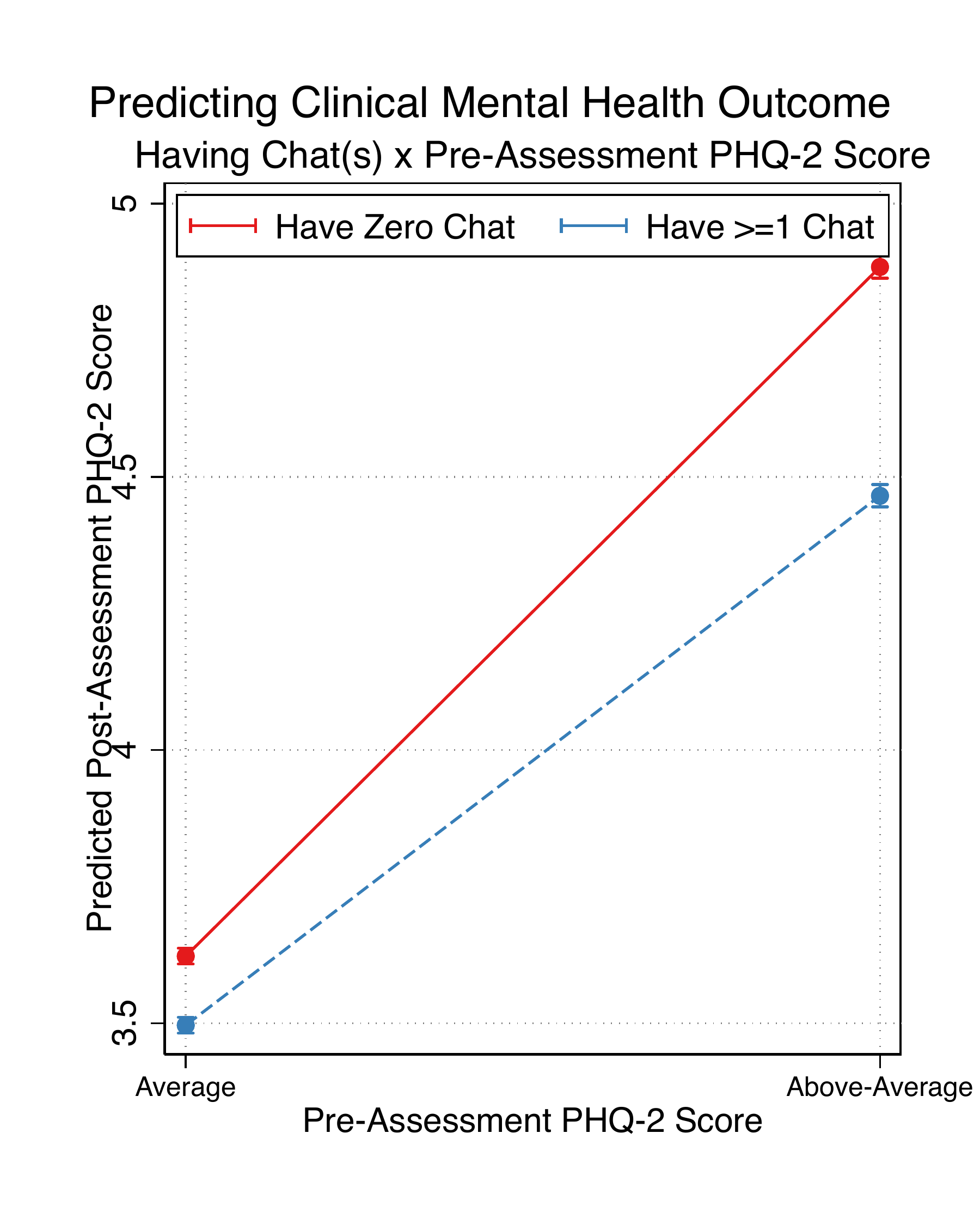}
     \includegraphics[scale=.2]{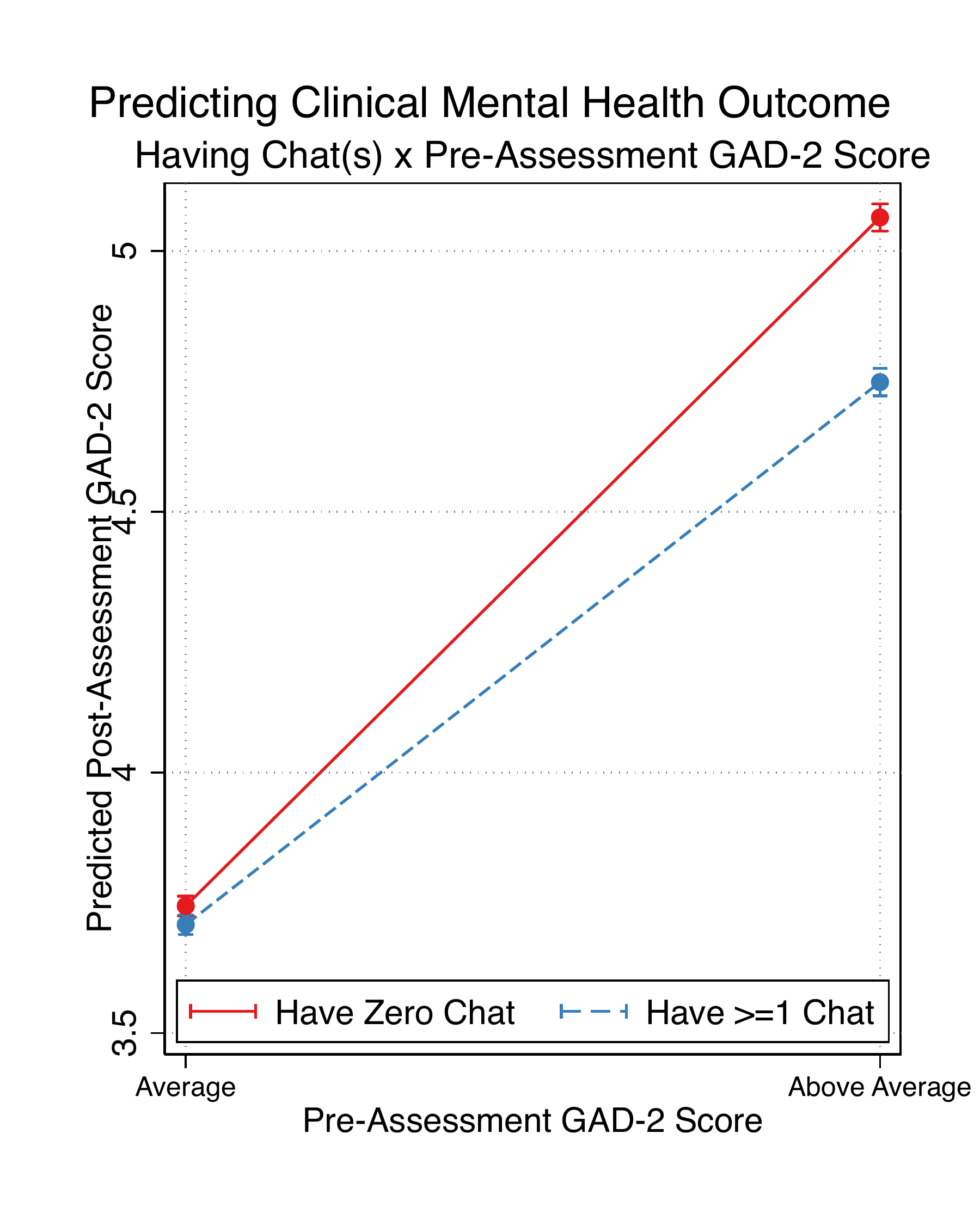}
  \caption{Marginal plot that illustrates the interaction effect between treatment and pre-assessment score, for support-seekers who have average vs. above-average initial symptom severity. Note that a higher score indicates \textit{worse} symptoms (i.e. greater symptom severity).}
    \label{/MH-chatXpre}
    \vspace{-1mm}
\end{figure}

Table \ref{RQ1-results} shows the results for RQ1, the effects of chats on changes in mental health. In terms of control variables, as expected, both depression ($\beta$ =.67, p<0.001) and anxiety ( $\beta$ =.68, p<0.001) were very consistent over the several weeks between pre- and post- assessments. In addition, support-seekers who had more on-platform conversations prior to the current observation period had lower post-assessment depression ($\beta$ =-.05, p<0.001) and anxiety ($\beta$ =-.07, p<0.001) scores.

More importantly, the results show that the treatment -- participation in support-chats during the observation period -- was associated with small but highly reliable improvements in mental health. Participation in support-chats was associated with improvements in symptoms of both depression ($\beta$ =-0.126, p < 0.001) anxiety, although to a lesser degree ($\beta$ = -0.036, p < 0.01). In addition, the Treatment $\times$ pre-assessment mental health interactions show the benefits of participating in the support chats were greater for those who initially had worse mental health for both depression ($\beta =-.155, p<.001$) and anxiety ($\beta -.145, p<.001$) . As illustrated in Figure ~\ref{/MH-chatXpre}, the dotted blue lines consistently fall beneath the solid red ones. This indicates that support-seekers who participate in chats consistently exhibit fewer symptoms of depression (PHQ-2) and anxiety (GAD-2) (indicative of improved mental health) than those who do not participate, regardless of their initial mental health status. In addition, the slopes of the lines differ, with the difference between those participating in chats versus not is greater when the pre-assessment measure of mental symptoms is high (i.e., 1) rather than low (0). This suggests that individuals with above-average (or major) depression or anxiety derive more benefit from engaging in OMHCs than those with average or mild symptoms.

\section{RQ2. What characteristics of the support conversations affect clinical mental health outcomes?}

\subsection{Methods}
To answer RQ2, we also adopted a regression analysis to study how therapeutic techniques that occurred during support conversations affected changes in support-seekers’ mental health. This research involved three steps.

\begin{enumerate}
\item Firstly, we collected consecutive PHQ-2 and GAD-2 assessments from support-seekers and arranged them in pairs, creating an observational period. 
\item For support-seekers who had a support chat during the observational period, we measured factors that the prior literature suggests should be associated with changes in support-seekers mental health. These include characteristics of the support providers and support-seekers before the observation period started and characteristics of the support conversations that occurred during the observations period. 
\item We used random-effects regression models to predict how the providers', seekers', and chat characteristics predict changes in mental health assessments (i.e., post-assessment controlling for pre-assessment).
\end{enumerate}

\textcolor{black}{For each observation period in which a support-seeker has at least one conversation, we used seven independent variables as listed below in 5.1.1 and 5.1.2.}
\subsubsection{Support-provider characteristics} We measured five characteristics of the support providers with whom the support-seeker chats during the observation period:
    \begin{itemize}
        \item \textbf{Chats with repeat support providers}: capturing the percentage of chats the support-seeker had with a support-provider they previously interacted with before the time of pre-assessment. This tracks the relationship between support-seekers and support providers, a crucial factor in counseling processes as suggested by previous studies \cite{Ackerman2003-ib, Wampold2023-nx}.
        \item \textbf{Support-providers' age}: a continuous variable representing the average age of the support providers interacted with by the support-seeker.
        \item \textbf{Support-providers' experience in chats}: a variable measuring the average total number of chats these support providers have engaged in from their registration until the time of pre-assessment. 
        \item \textbf{Support-providers’ tenure}: time since the support provider registered on the online support platform.
        \item \textbf{Support-providers' training modules}: the average number of awards for support providers for completing additional training modules by the platform. 

    \end{itemize}
\subsubsection{Support-seeker characteristics} We measured two characteristics of the support-seeker before the start of an observation period. 
    \begin{itemize} 
        \item\textbf{Support-seekers’ pre-assessment mental health scores} (i.e., PHQ-2 or GAD-2): By including pre-assessment score, this lagged dependent variable model effectively examines how the independent variables predict changes in mental health status \cite{cohen2013statistical}.
        \item \textbf{Support-seekers’ tenure}:  As a measure of the support-seekers' experience with the platform, we included time since the support-seeker registered on the online support platform. Because this measure had a  long tail, we used the log transformation.  

        \end{itemize}
 \subsection{Chat characteristics.} 
  We concatenated all the communication the support-seeker had with one or more providers during an observation period and measured the following attributes of these conversations:      
     \begin{itemize} 
        \item \textbf{Total conversation turns}: The number of conversational turns during an observation period is a measure of the length of the conversation. We include it to gauge the "dosage effect" of peer-support chats on mental health. Just as the analysis for  RQ1 showed that having a chat was more strongly associated with improvements in mental health for those who initially had worse mental health at the beginning of the observation period,  we included  interaction term between total turns and pre-assessment mental health . 
        \item  \textbf{Sentiment in support-seekers' language}: The positivity or negativity of the support-seekers language during the conversations may be an indicator of their mental health. In addition, previous research suggests the positivity or negativity of support-seekers' language may influence how support providers respond to them. For example, providers may offer more emotional support in response to seekers using emotionally negative language \cite{Wang2012-ny}. To measure sentiment, we adopted the off-the-shelf VADER classifier  from Hutto et al.’s \cite{hutto2014vader}. VADER is a rule-based model for sentiment analysis that performs exceptionally well for social media text, which makes it particularly relevant for our study’s case. It outperforms human raters in social media datasets with an F1 score of 0.96. We applied the model to the support-seekers' language during the chats. The VADER classifier outputs an inclusive three-class label (negative, neural, and positive), and we included the negative and positive labels as predictors.
        \item \textbf{Support providers' therapeutic language}: We used natural-language processing techniques (NLP) to measure the support providers' language.  We focused on Motivational Interviewing behaviors and empathy, two kinds of therapeutic techniques that have been widely shown to be effective in traditional counseling contexts and are used for training volunteers in online support platforms’. 
     \begin{itemize} 
       \item  \textbf{Motivational Interviewing behaviors}: We measured seven Motivational Interviewing behaviors from the MITI 4.2 manual \cite{Moyers2016-kx}, including asking questions, offering affirmations, and reflecting back the seekers' thoughts,  using the classifiers developed by Shah et al \cite{Shah2022-pj}. The MITI code classifiers are transformer-based models (i.e., BERT) that were pre-trained on 120 million chat messages on the same platform as our study and fine-tuned based on \textcolor{black}{manually annotated datasets. These manually annotated data consists of 14797 utterances from 734 conversations annotated with 17 Motivational Interviewing behaviors adapted for the online nature of the platform of study by researchers of \cite{Shah2022-pj}. Each model takes in a support-provider’s utterance as input, as well as the immediately previous utterance, and outputs a binary label indicating whether the current utterance contains the corresponding MITI code.} Descriptions of these measures, including definitions, examples, and classification accuracy are shown in Table \ref{table:miti-empathy-codes}.
       \item \textbf{Empathy}: We measured three components of empathy (emotional reaction, exploration, and interpretation) based on classifiers developed by Sharma et al. \cite{Sharma2020-il}. Descriptions of these measures, including definitions, examples, and classification accuracy are shown in Table \ref{table:miti-empathy-codes}. The classifiers include two independently pre-trained RoBERTa-based encoders that accept one turn of clients’ and support providers’ messages separately and identify empathy shown in the support providers’ responses. Following Sharma et al.’s settings, we fine-tuned the models on a public Reddit dataset \cite{Sharma2018-iv} that contains many mental health communities. As reported in their paper, the models have F1 scores ranging from 62.60 to 74.46 in terms of identifying three kinds of empathy on the Reddit dataset.
    \end{itemize} 
    \end{itemize}

\begin{table}
\centering
\def\arraystretch{1.3}

\begin{tabular}{|>{\hspace{0pt}}m{0.08\linewidth}|>{\hspace{0pt}}m{0.12\linewidth}|>{\hspace{0pt}}m{0.332\linewidth}|>{\hspace{0pt}}m{0.229\linewidth}|>{\hspace{0pt}}m{0.13\linewidth}|} 
\hline
\multicolumn{2}{|>{\centering\hspace{0pt}}m{0.2\linewidth}|}{\textbf{Therapeutic Techniques}} & \centering\textbf{Description}& \multicolumn{1}{>{\centering\hspace{0pt}}m{0.229\linewidth}|}{\textbf{Example}}& \multicolumn{1}{>{\centering\arraybackslash\hspace{0pt}}m{0.13\linewidth}|}{\textbf{Classifiers’}\par{}\textbf{F1 Score}}\\ 
\hline
\multirow{7}{\linewidth}{\hspace{0pt}MI} & Giving \par{}Information& Support-provider gives general information or feedback in a neutral tone. & \textit{“This website has a lot of materials with coping skills for anxiety if you would like to do more research.”}& 0.574 \\ 
\cline{2-5}
& Question& Support-provider asks a question that may be either open (leaving space for support-seekers’ response) or closed (implies a short or restricted answer).& \textit{“Hello, what would you like to talk about?”}\par{}\textit{“How old were you when you had your first drink?”}& 0.927 \\ 
\cline{2-5}
& Reflection~ & Support-provider makes a reflective listening statement, capturing and returning something the support-seeker has just said. & \textit{“It seems like you are having a hard time figuring out the way forward.”}& 0.692 \\ 
\cline{2-5}
& Affirm& Support-provider accentuates something positive about the support-seeker, such as their strengths, intentions, or behavior.& \textit{“It’s important to you to be a good parent, just like your folks were for you.”} & 0.624 \\ 
\cline{2-5}
& Seeking\par{}Collaboration~& Support-provider attempts to share power or acknowledge the client’s expertise (e.g. support-provider seeks consensus with the client regarding tasks) & \textit{“Would it be alright if we spend some time discussing the standards for consuming alcohol during pregnancy.”} & 0.500 \\ 
\cline{2-5}
& Emphasizing\par{}Autonomy & Support-provider focuses on support-seekers’ sense of control, freedom, and ability to decide their actions. & \textit{“You’re the one who knows yourself best here. What do you think ought to be on this treatment plan?”} & 0.785 \\ 
\cline{2-5}
& Persuade with\par{}Permission& Support-provider emphasizes collaboration or autonomy support while persuading. Permission is present when e.g. client asks for opinion or support-provider asks for permission. & \textit{“I have some information about your risk of problem drinking and I wonder if I can share with you.”}& 0.770 \\ 
\hhline{|=====|}
\multirow{3}{\linewidth}{\hspace{0pt}Empathy} & Emotional\par{}Reactions& Support-provider expresses warmth, compassion, and concern.~ & \textit{“I’m so sorry that you experienced this”} & 0.745 \\ 
\cline{2-5}
& Interpretations & Support-provider communicates that they understand the feelings/experiences of the support-seeker, and may attempt to describe or relate. & \textit{“Honestly I’ve been through the same thing. My parents got divorced when I was 18...I understand why you would be guarded.”} & 0.626 \\ 
\cline{2-5}
& Exploration & Support-provider attempts to explore, probe, or describe the feelings/experiences not explicitly expressed by the support-seeker. & \textit{“Do you feel like you have a short temper at the moment or that you’re more snappy than usual”} & 0.726 \\
\hline
\end{tabular}
\caption{Description, examples, and classifier scores for our study's seven MI (MITI 4.2) codes \cite{Moyers2016-kx} and three kinds of empathy \cite{Sharma2020-il}. Classifiers are from \cite{Shah2022-pj} (MI) and \cite{Sharma2020-il} (empathy).}
\label{table:miti-empathy-codes}
\end{table}

\subsubsection{Regression Analysis}
\begin{figure}[]
    \centering
    \includegraphics[scale=.55]{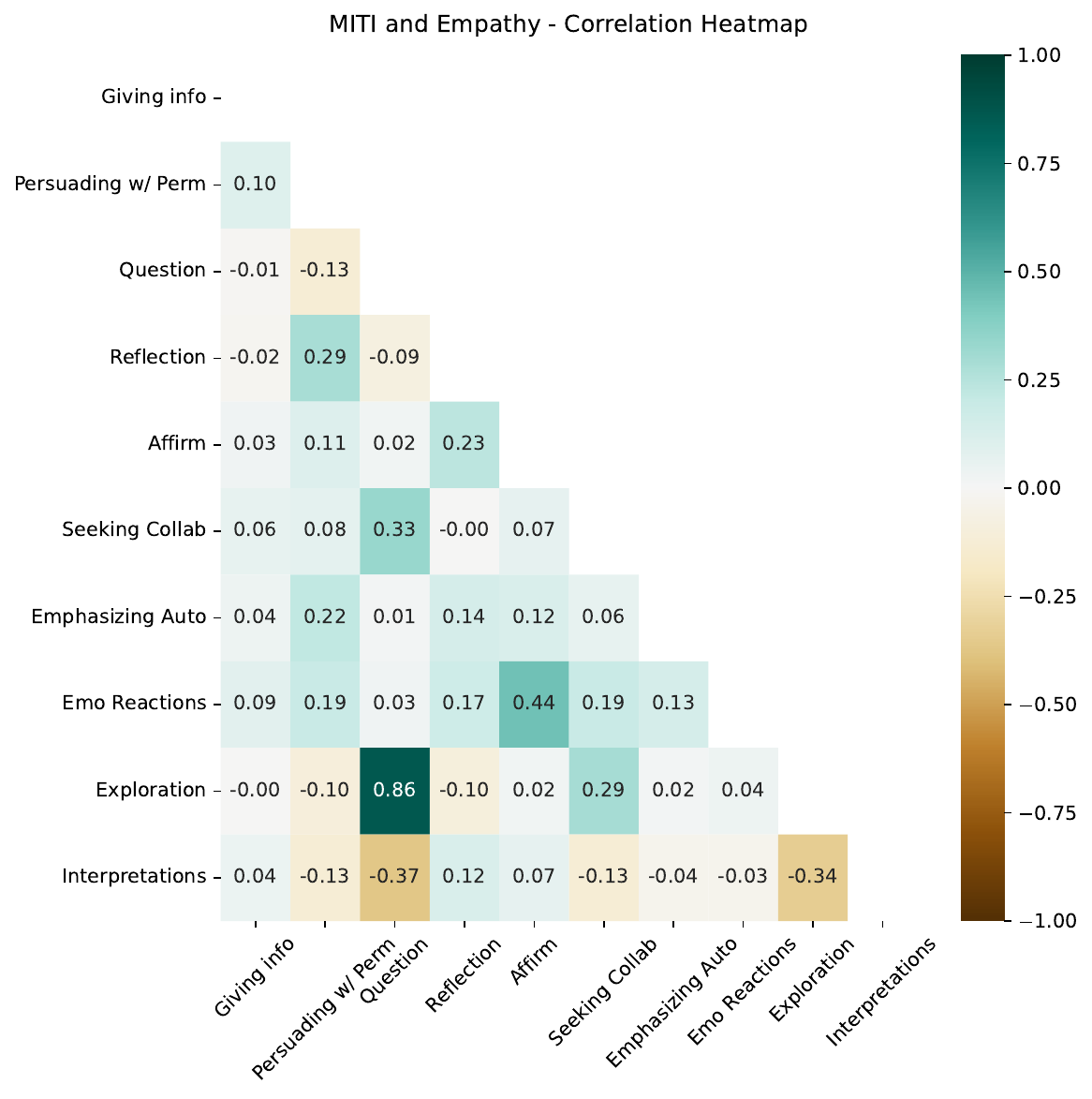}
  \caption{Correlation heat map between all MI codes (Giving info, Persuading, Question, Reflection, Affirm, Seeking Collab, Emphasizing Auto) and Empathy codes (EmoReactions, Exploration, Interpretation).}
    \label{figure:mi-empathy-corr}
\end{figure}

In order to understand how support-provider characteristics, support-seeker characteristics, and characteristics of the conversations predict changes in mental health, we used a similar random-effect model specification as in RQ1 with additional predictor and control variables. However, because the RQ2 analysis contains many more predictor variables and all predictors are continuous rather than binary, we cannot use propensity score matching as a pre-processing step. Results are shown in Table \ref{RQ2-table}. Because some of the measures of MI behaviors are highly correlated with the measures of empathy, resulting in multi-co-linearity, we conducted separate analyses including MI interviewing behaviors and empathy as predictors. The correlation between MI codes and Empathy codes are shown in Figure \ref{figure:mi-empathy-corr}. 

\subsection{Results}
\begin{table}[!ht]
    \centering
    \resizebox{\textwidth}{!}{
    \begin{tabular}{|l|l|l|l|l|l|l|l|l|} 
    \toprule
        Dependent Variables  & \multicolumn{4}{c|}{\textbf{PHQ-2 ~ ~ ~}}       & \multicolumn{4}{c|}{\textbf{GAD-2 ~ ~ ~}}        \\
        Model No.   & (1) MITI  & ~& (2) Empathy & ~& (3) MITI  & ~& (4) Empathy & ~ \\
        Independent Variables& $\beta$ & Robust SE & $\beta$& Robust SE & $\beta$       & Robust SE & $\beta$& Robust SE  \\ 
        \midrule
        Constant & 3.526*** & 0.009 & 3.526*** & 0.009 & 3.737*** & 0.011 & 3.737*** & 0.011 \\     
Support-provider's age & -0.036*** & 0.009 & -0.039*** & 0.009 & -0.019 & 0.011 & -0.016 & 0.011 \\ 
        Support-provider’s chats (log) & -0.001 & 0.009 & 0.003 & 0.009 & -0.012 & 0.011 & -0.011 & 0.011 \\ 
        Support-provider’s training modules (log) & 0.017 & 0.009 & 0.011 & 0.009 & 0.037*** & 0.011 & 0.030** & 0.011 \\ 
        \% chats with same supporter  & 0.015 & 0.009 & 0.012 & 0.009 & 0 & 0.011 & -0.002 & 0.011 \\ \hline
        Support-seeker’s  time on platform (log) & -0.005 & 0.009 & -0.007 & 0.009 & -0.01 & 0.011 & -0.01 & 0.011 \\ 
        Pre-assessment mental health score (MHS) & 0.446*** & 0.005 & 0.446*** & 0.005 & 0.460*** & 0.007 & 0.460*** & 0.007 \\ \hline
        Total turns (log) & -0.023* & 0.009 & -0.021* & 0.01 & 0.012 & 0.012 & 0.023 & 0.012 \\ 
        Total turns x MHS & -0.071*** & 0.005 & -0.071*** & 0.005 & -0.056*** & 0.006 & -0.056*** & 0.006 \\ 
        Support-seeker’s negative sentiment & 0.208*** & 0.009 & 0.208*** & 0.009 & 0.193*** & 0.012 & 0.193*** & 0.011 \\ 
        Support-seeker’s positive sentiment & -0.024** & 0.009 & -0.028** & 0.009 & -0.011 & 0.011 & -0.018 & 0.011 \\ 
        Provider gives information & 0.019* & 0.008 & ~ & ~ & 0.017 & 0.01 & ~ & ~ \\ 
        Provider persuades with permission & 0.022** & 0.009 & ~ & ~ & 0.033** & 0.011 & ~ & ~ \\ 
        Provider questions & 0.011 & 0.009 & ~ & ~ & 0.005 & 0.012 & ~ & ~ \\ 
        Provider reflection & -0.022* & 0.009 & ~ & ~ & -0.008 & 0.011 & ~ & ~ \\ 
        Provider affirmation & -0.007 & 0.008 & ~ & ~ & 0.007 & 0.01 & ~ & ~ \\ 
        Provider seek collaboration & 0.016* & 0.008 & ~ & ~ & 0.001 & 0.011 & ~ & ~ \\ 
        Provider emphasize autonomy & -0.020* & 0.008 & ~ & ~ & 0.005 & 0.01 & ~ & ~ \\ \hline
        Provider emotional reaction & ~ & ~ & 0.028*** & 0.009 & ~ & ~ & 0.048*** & 0.011 \\ 
        Provider exploration & ~ & ~ & 0.014 & 0.009 & ~ & ~ & 0.002 & 0.011 \\ 
        Provider interpretation & ~ & ~ & 0.012 & 0.009 & ~ & ~ & 0.008 & 0.011 \\ 
        \midrule
        N\_respondents & 25536 & ~ & 25536 & ~ & 15241 & ~ & 15241 & ~ \\
        N\_obs & 41751 & ~ & 41751 & ~ & 25689 & ~ & 25689 & ~ \\ 
        R-squared  overall & 0.316 & ~ & 0.316 & ~ & 0.344 & ~ & 0.344  & ~\\ 
    \bottomrule
    \end{tabular}
    }
    \caption{Random effects regression predicting support-seekers’ post-assessment PHQ-2 and GAD-2 mental health score, from characters of support providers and support-seekers and characteristics of the support-chats. 
    *: p<0.05, **: p<0.01, ***: p<0.001}
    \label{RQ2-table}
\end{table}
\textcolor{black}{Table \ref{RQ2-table} shows that some characteristics of the support-providers were associated with statistically significant changes in seekers' mental health. We note that our findings generally show small R-squared values, though; we further discuss this in Section 6.1.1.} We had expected that chatting with providers with more experience would be associated with improvements in mental health, but this expectations was only partially confirmed. In particular, talking with older support-providers was associated with reductions in depression ($\beta  = -.036, p<.001$). This may be because the relatively young support-seekers on this site have better support experiences when they talk to others older than themselves with richer life experiences \cite{Fang2022-gc}. However, there were not significant effects for anxiety. Surprisingly, support-providers' experience on the site, including past chatting experience and having repeated interactions with the support seeker during the observation period were not associated with changes in either depression or anxiety (p>0.05). In fact, talking with support-providers who completed more training modules on the site was associated with increases in anxiety ($\beta = .037, p < .001$). We discuss finding more in Section 6 Discussion, and given correlational evidence we cannot determine the cause of this reversal.\label{table:psm-balance}

We now turn to the associations of changes in mental health with support-seekers' characteristics. Support-seekers with a unit higher pre-assessment PHQ-2 score had higher post-assessment scores as well ($\beta=0.446, p<0.001$). Similarly, those with with a unit higher pre-assessment GAD-2 anxiety score reported higher anxiety during post-assessment ($\beta$=0.46, p<0.001). The sentiment of support-seekers in their messages predicted changes in their mental health. In particular, support-seekers who expressed more negative sentiment in conversations during an observation period reported worse depression ($\beta =0.208, p<0.001$) and anxiety ($\beta = 0.193, p<0.001$) on their post-assessments. In addition, the associations with positive sentiment followed a similar but weaker pattern. Support-seekers who expressed more positive sentiment in conversations during an observation period reported slightly lower depression ($\beta = -0.024, p<0.01$), but no change in their levels of anxiety ($\beta = -0.001, p>.05$) on their post-assessments.

\begin{figure}[!h]
    \centering
    \includegraphics[scale=.2]{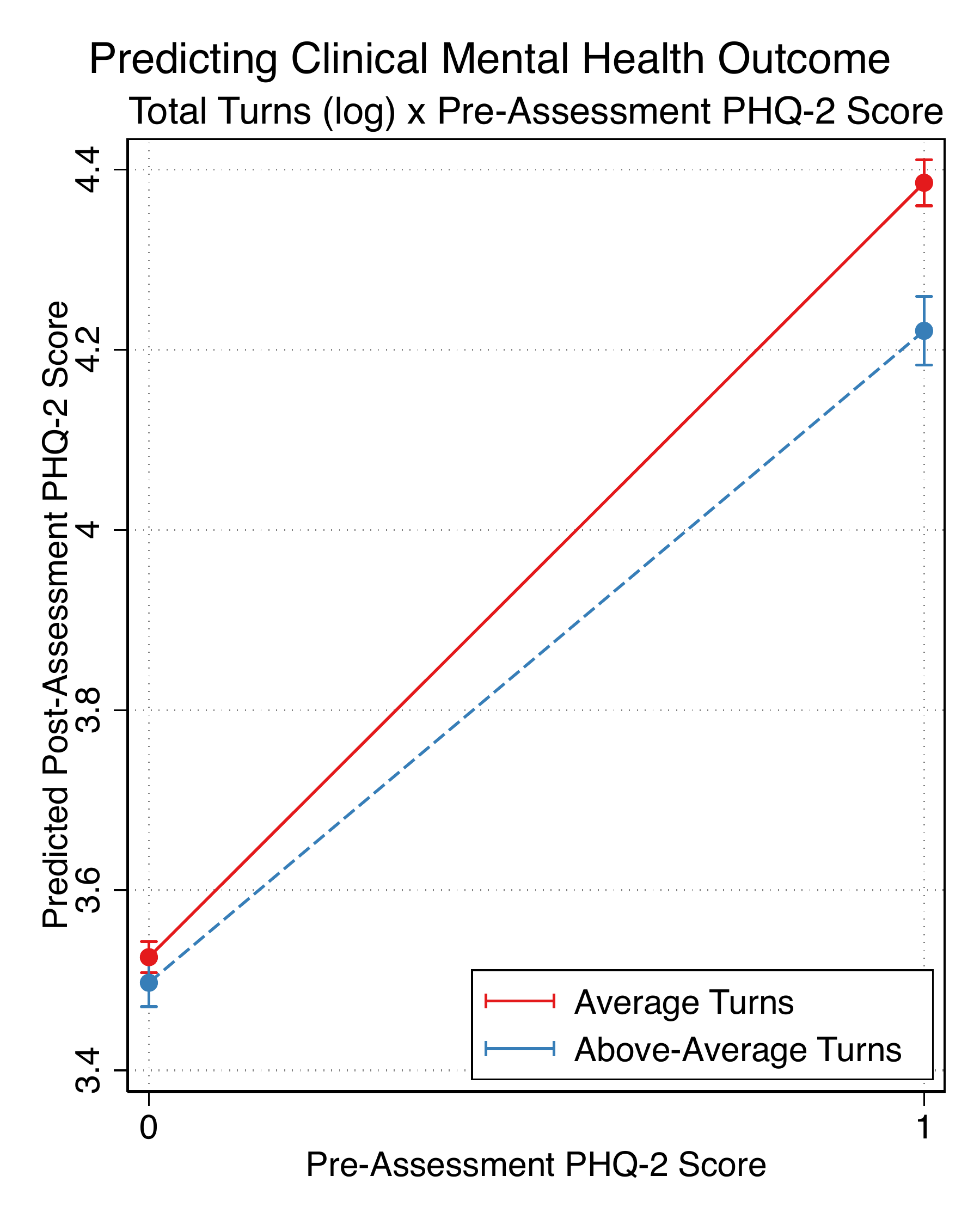}
     \includegraphics[scale=.2]{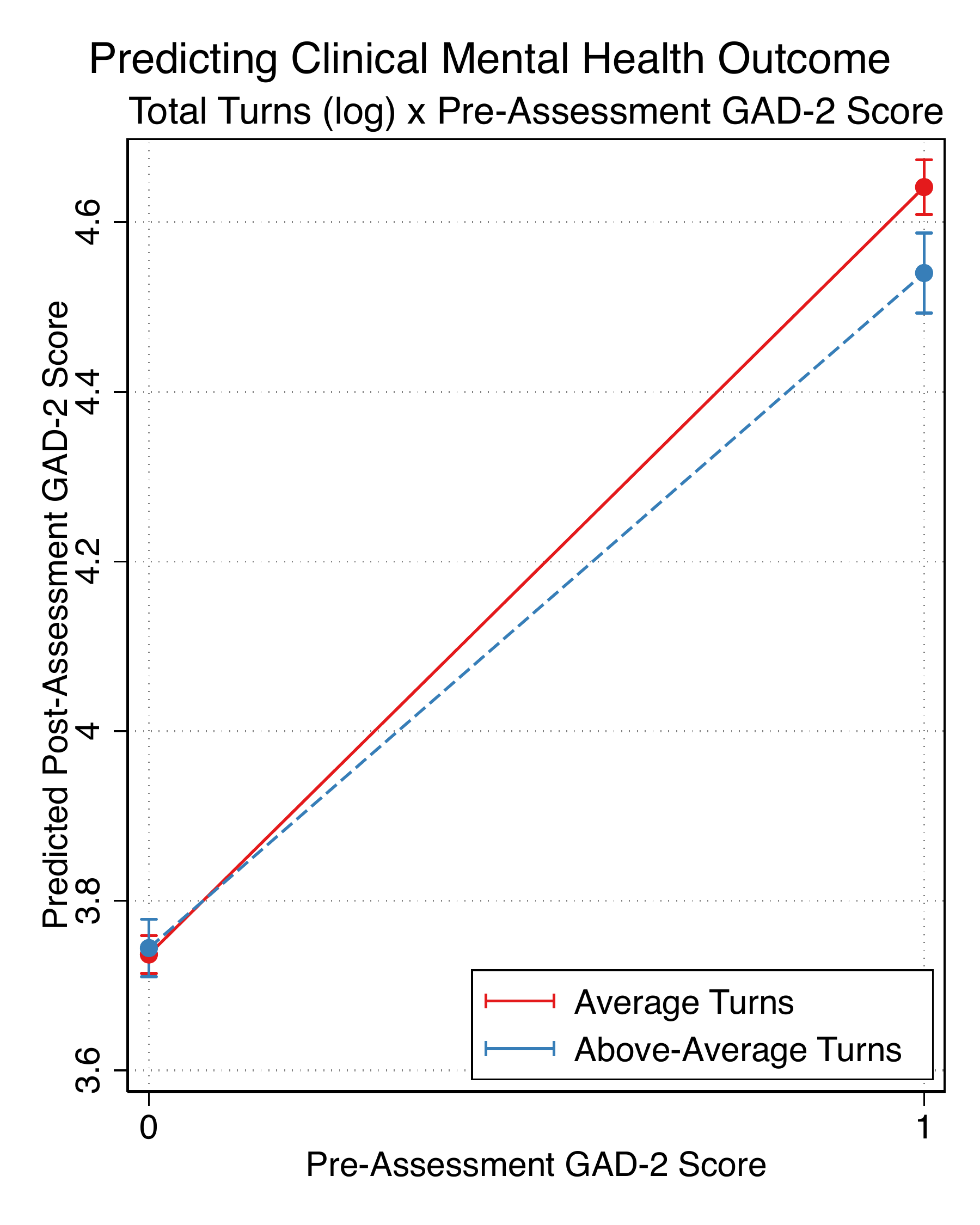}
  \caption{Marginal plot that illustrates the interaction effect between total conversation turns and support-seekers' pre-assessment scores, comparing effects for those with average versus above-average initial symptom severity.}
    \label{/MH-turnXpre}
\end{figure}

We now review the associations of changes in mental health with characteristics of the support conversations. Our analysis for RQ1 demonstrated reliable improvements in depression and anxiety associated with participating in online peer support conversations, with the improvements larger for those who initially had worse mental health. The RQ2 analysis shows very similar results among those who participated in the support-chats. We consider the length of the support chats as a "dosage" of a support conversation; people who participated in longer chats with more turns have slightly improved mental health for depression ($\beta=-.023, p <.05$) but not anxiety ($\beta=.012, p >.05$), with larger effect size for those who initially showed worse symptoms. Having longer chats was more strongly associated with improvements in mental health for those who initially had worse depression ($\beta=-.071, p <.001$) and for those who initially had worse anxiety ($\beta=-.056, p <.001$). These interaction effects are illustrated in Figure \ref{/MH-turnXpre}.

In terms of the content of the support-conversations, our RQ2 results suggest that some MI and empathy behaviors positively predict improvements in mental health, but others are actually associated with declines in mental health status. Out of the seven MI behaviors, Reflections and Emphasizing Autonomy were significantly associated with improvements in depression symptoms. Specifically, when people received one standard deviation more Reflection and Emphasizing Autonomy during an observation period their PH-2 depression scores improved by 0.022 ($p<.05$) and 0.02 ($p<.05$) units respectively. In contrast, when they received a standard deviation more Information, Persuade with Permission, and Seeking Collaboration their PH-2 depression outcomes got worse (p<0.05). \textcolor{black}{However, we note that Seeking Collaboration and Giving Information both have only low to moderate accuracy in our MI classifiers (0.5 and 0.574, respectively), so both of these significant results may need to be interpreted with some caution.} In terms of empathy codes (Emotional Reaction, Exploration, Interpretation), only Emotional Reaction had significant association with changes in mental health. Inconsistent with much research on the common factors that seem responsible for success in conventional psychotherapy \cite{Norcross2019-xh}, receiving Emotional Reaction during the observation period was associated with a small, but statistically reliable increase of 0.028 ($p<.001$) points on their PHQ-2 score.

In terms of anxiety symptoms, a standard deviation increase of Persuade with Permission similarly led to worsening of anxiety symptoms ($\beta$=0.033, p<0.001). Notably, there were no MITI codes that led to statistically reliable  improvements in  anxiety. Similar to our PHQ-2 analysis, support-seekers had a worse score on GAD-2 when provided Emotional Reaction. One standard deviation increase in receiving Emotional Reaction during the observation period was associated with a small, but statistically reliable increase of 0.048 ($p<.001$) point increase in their GAD-2 score.

\section{Discussion}
\textcolor{black}{Our work strongly suggests that online peer support chats are useful for improving clinical depression and anxiety outcomes, with a relatively small effect size.} Additionally, we found effective MI and empathy skills like Reflection and Emphasizing Autonomy that contribute to this clinical effectiveness. However, several techniques including Seek Collaboration, Persuading (with Permission), Give Information, and Emotional Reaction were associated with worsening of mental health outcomes. Our work has important implications for understanding how online support chats affect people's mental health symptoms as well as the specific methods contributing to many OMHCs' training that are effective when used by online volunteer support providers.

\subsection{Effectiveness of Online Peer Support Chats}
After matching on factors likely to be associated with mental health and willingness to engage in support chats, support-seekers who chat on the platform report less severe symptoms of depression and anxiety compared to the matched seekers who did not engage, with the change highly reliable for both depression and anxiety (p $<$ .001). Moreover, support-seekers who initially had more severe depression and anxiety showed greater improvement compared to those with milder initial symptoms. Furthermore, our analysis for RQ2 shows a parallel result, with the dosage of the chats (i.e., more turns) being associated with improved mental health, and especially so among support-seekers' who initially had worse depression and anxiety. Thus, findings for both RQ1 and RQ2 suggest that online support chats are more effective the more distressed a support-seeker is. Our study contributes a clinical perspective on existing literature \cite{Wang2023-um, Doherty2012-da, Althoff2016-dt} about the general effectiveness of online communities for mental health outcomes. Although the effect sizes are small, these results are consistent with the hypothesis that participating in peer-to-peer support-conversations with lightly trained volunteers improves people's mental health.  

Although the results suggest that peer-to-peer support can improve mental health, our research did not conclusively identify the chat techniques and characteristics that account for this success. For example, we found that no MI techniques had significant results on improving anxiety outcome. Although further research is required to conclusively identify the cause, one proposed reason for the insignificance of all MI codes for anxiety may be that generally engaging in a support chat (or participating in an online health forum) is itself beneficial for one’s anxiety, given that as reviewed personal disclosure and emotional expression in general is healthy for both emotional and physical well-being \cite{Kennedy-Moore2001-ul, Pennebaker2012-dc}. As Pennebaker's research suggests \cite{Pennebaker2012-dc}, it may be that merely engaging in the conversations and having an opportunity to talk about ones problems and express ones emotions is all that is necessary. According to this view, role of the support-provider is to provide an audience to encourage fruitful self-expression rather than to act as an amateur therapist. Additionally, prior work has found that distraction from negative emotions leads to improved mood. This distraction may occur through just casually chatting through online social channels like social media platforms and OMHCs , where people also engage in casual chatter not about any mental health issue \cite{Dalebroux2008-qv, Shah2022-pj}. Given this, our findings may indicate that (in addition to the techniques of support providers that are effective in this context) simply having the space to express in the first place can relieve mental distress. Further work may find conclusive explanations for why there are positive effects of peer support chats on anxiety outcomes.

We also found other counter-intuitive results such as the MI technique of Seek Collaboration predicting worse depression outcomes. Emotional Reaction was also related to a significant increase (i.e. worsening mental health) in both PHQ-2 and GAD-2 score. Although our analysis methods do not allow us to conclusively find direction of causality, one possible explanation for this finding is that support-seekers who express more negative or harmful sentiments elicit certain support-provider techniques, such as emotional reaction. Our analysis method is limited in the fact that we labeled support-providers' utterances independently of the context of the support chat and support-seekers' messages. However, in reality, we expect that support-providers' chat behaviors largely depend on the support-seekers' messages. Thus, although our analysis results suggest these MI-consistent techniques predict worse mental health outcomes, it is the support-seekers' mental health state and chat dynamics that elicit these behaviors. 

Despite finding some inconsistent results that need further research to dissect, we did find several effective behaviors of support providers that align with prior work in MI and therapy generally in face-to-face contexts. Reflection and Emphasizing Autonomy MI techniques have both been found in MI research to be consistent with therapeutic goals and lead to better mental health changes \cite{RN9261, RN9262}. Our work likewise found that both of these skills improved depression outcomes in the online setting as well. Research in therapy also concludes that people commonly find emotional relief even without added substantive or personalized response, such as through expressing personal struggles through writing, involuntary expression (e.g. crying), or even through talking to strangers \cite{Pennebaker2012-dc}. Client-led sessions through using skills like Reflection, where the support-provider does not necessarily add substantive comment but instead just reiterates what the support-seeker has already said, may be effective in the online context as simply providing the space for emotional expression itself relieves distress. The effectiveness of Emphasizing Autonomy also follows work in traditional offline therapy that has found respect for the autonomy of a client leads towards better therapy outcomes, due to reasons such as providing more motivation for the client and building a better client-therapist relationship \cite{fisher2008informed,scheel2011client}

We also found MI techniques Giving Information, Persuasion (with Permission), and Seeking Collaboration related to worse depression outcomes. Prior work on MI \cite{Magill2014-qu, magill2018meta} has found that "MI-inconsistent" techniques, which include aggressive behaviors like confronting, unsolicited advice, or persuading the client, are harmful to health. Our findings partially align with this past work, as we found a significant rise of 0.022 in PHQ-2 score with increased Persuasion (although with permission, in our case). Although our coding for Persuasion is conditioned on permission being present, we note that permission can take different complex forms from either the support-seeker or provider's side, and the classifier for seeking permission that is used in the classifier from prior work \cite{Shah2022-pj} that we use in our study has lower accuracy at 0.5. We also found that Giving Information and Seeking Collaboration both had harmful effects. \textcolor{black}{However, an important note is that both these MI codes also have lower classification accuracy than our other MI codes (see Table 4), so these findings may not be as reliable. Still, one possible reason that providing information could plausibly lead to harmful effects is if information provided was unsolicited. Support-seekers may be seeking out emotional rather than informational support when using an online peer support service; however, this would need further exploration given that prior work in the context of an online breast cancer support community found people seeking emotional support are satisfied with either emotional or informational support \cite{Vlahovic2014-ws}.} 

\subsubsection{\textcolor{black}{Low Effect Size of Support-Provider Techniques}}

\textcolor{black}{We also specifically discuss the clinical significance but low R-squared values in our results, indicating a low level of the variations in our assessment scores can be attributed to our independent variables. In addition to low explanatory power, our analysis also overall showed low effect sizes indicating statistically significant but small improvements. There are potentially several reasons for these findings. Firstly, at the simplest level, our findings of small effect sizes for support-provider behaviors indicate that what a support-provider simply does not have a particularly strong effect on a support-seekers' mental health status. This may be due to support-providers having minimal training as volunteer counselors, and thus are not properly trained to provide support in effective ways. As we have also discussed, simply having the space for expression may be useful in itself rather than a support-providers' behaviors specifically. Secondly, our analysis is limited by the ways we counted support-provider techniques being used, rather than if they were being used in relevant and helpful ways. Our analysis, taking from classifiers in \cite{Shah2022-pj}, included simply labeling MI codes when used rather than taking in context from an entire conversation or how the support-seeker may have elicited a certain support behavior from a counselor. In this way, our analysis may be lacking the context needed to understand \textit{how} MI and empathy codes were used and we thus limited the effect sizes we could see. It is highly plausible that support-provider techniques being used in the right ways would have a greater improvement effect. We note that this is important to explore in future work not just for our study but in the field generally; analyzing text conversations for how people exchange support requires complex understanding of whether these techniques and conversations are being used in productive ways given topic, support-seeker chats and responses, support-providers' experience and language use, and more. Lastly, we note there are many other significant variables that contribute to how support-seekers' health improves apart from just participating in our study's online mental health community platform. It is not uncommon to encounter low R-squared values in online community research and in studying social phenomena generally \cite{Wooldridge2015-kd}, given the huge number of variables that influence user behaviors on these platforms. For example, we unfortunately have no insight into support-seekers' behaviors regarding their mental healthcare off of the platform. Our analysis did not capture relevant variables that presumably can explain assessment outcomes (e.g. other treatment-seeking, lifestyles). Additionally, we acknowledge some limitations of our platform-specific study that could have influenced the explanatory power, such as chats on the platform being generally short (~23 minutes) and using the shortened versions of PHQ and GAD assessments.}

\subsection{Designing Training for Online Support Platforms}

Our study generally contributes to the design of training for digital support platforms. In particular, our study identified how the therapeutic techniques that guide training in online support platforms, which are widely established in therapy and social work contexts, can significantly help or harm people. We found evidence that several established therapeutic techniques translate to effective clinical outcomes in the online context; for example, our findings suggest that focusing training on client-led techniques like emphasizing a support-seekers' autonomy and reflecting back their own stated feelings has effective outcomes, while behaviors that could be thought of as more support-provider-led (persuading, giving information) are harmful for people's mental health outcomes. 

We found insignificant results for some behaviors that would presumably contribute to effective care (e.g. affirmation). \textcolor{black}{As we have mentioned, the ways that our work classifies MI techniques and empathy  are context-blind; we only label an utterance as containing a technique or not rather than  whether the technique is being used \textit{appropriately}. This is a problem prevalent in this research area generally, where online mental health interventions aim to increase the frequency at which people use support techniques with minimal consideration of \textit{when} or \textit{how} these techniques should be used. For example, Sharma et al's HAILEY system, which helps volunteer supporters write empathetic responses, doesn't focus on when to offer empathy \cite{sharma23-Empathy}. This focus on  quantity rather than appropriateness may contribute both  to some support-provider behaviors showing insignificant results and low effect size even for statistically significant findings.} 

\textcolor{black}{Additionally, some of our findings seem counterintuitive, like providers' emotionally empathetic reactions predicting \textit{more} depression and anxiety or conversations with support-providers with more training modules predicting \textit{more} anxiety. It may be that unmeasured confounds account these surprising results. For example, providers' may offer more empathetic responses when talking to someone describing with worse  depression or anxiety. Similarly, the support-providers who feel less confident or perform poorly in support chats may take these additional training modules. Further analysis is needed to better understand the conditions under which these skills harm support-seekers and how to change online training programs to account for this. We also note that, as reviewed previously in Section 2.2, skills used in the online context versus traditional in-person care may differ in effect. Given that our existing frameworks such as MI are based in the traditional context, many general concepts of care that show in our analysis to be insignificant (or even harmful) may remain relevant in online support but appear differently when presented in text-based conversations. Alternatively, MI techniques and empathy may have unique behaviors when combined with behaviors that are specific to the online context (i.e. chit-chatting, demographic self-disclosure \cite{Fang2022-gc, Shah2022-pj}). There may be context-specific online behaviors which are not covered by existing frameworks; for example, effective techniques like casual chit-chatting that can distract from negative emotions may be helpful for mental health outcomes \cite{Dalebroux2008-qv, Shah2022-pj}.} We suggest that future work may study techniques within the online context specifically rather than just attempting to emulate face-to-face contexts; existing research suggests that emotions are expressed differently through text-based conversations due to reasons such as lack of audio/visual cues and asynchronicity, and trusting relationships are also different compared to in-person interactions \cite{Vermeulen2018-ah,Baker2011-ww, Elleven2004-nz, Richards2013-ef}

In general, our study is most relevant to the growing research interest in developing digital, self-guided, and interactive training to support online health communities \cite{Busse2021-op, Lee2020-qe, Posadzki2019-fo}. Digital interactive training environment can help to train novice support providers in using the non-specific clinical micro-skills and communication styles that are common across evidence-based therapies, such as Motivational Interviewing and Cognitive Behavioral Therapy (CBT). This type of training could provide scalable training to hundreds of thousands of volunteer support providers on online mental health platforms. Although we focus on mental health volunteers, a similar approach could be used to supplement the more formal education that medical practitioners receive. Although further work is needed to understand the exact skills that should guide online training, our analysis suggests promising effects of online support chats for depression and anxiety. Our analysis also found that (at least a subset of) existing frameworks of motivational interviewing and empathy do translate to beneficial effects for online support-seekers. Our work contributes to studying behavioral interventions and therapy strategies built for the online context, such as iCBT \cite{Chikersal2020-pf} and augments the findings of \cite{Shah2022-pj, Wang2023-um} in that outcome measures for measuring the effectiveness of these frameworks can be highly variable; some therapeutic behaviors may significantly affect proxies like support-seeker satisfaction but do not necessarily translate to clinical symptom improvement.

\subsection{Limitations}
Our work has the following limitations. First, we studied one peer support platform, and cannot guarantee our findings generalize to all other online platforms. Furthermore, our findings may not be applicable to the several OMHCs that do not include peer support chats or volunteer counselors, such as those that connect people to licensed therapists for teletherapy (e.g. BetterHelp.com) or provide support primarily through forum-based help (e.g. mentalhealthforum.net). Second, our analysis may have suffered from some limitations in measuring variables. Our analysis can only detect changes in mental health between any two PHQ-2 and GAD-2 questionnaires. Although our analysis attempted to identify the results of engaging in one or more peer support chats during that time, people can engage in support chats with several different support providers, engage in other platform activities, and more; as a result, it is difficult to isolate and assess the effects of having any given chat. Third, although our study uses longitudinal data on assessment score changes and uses propensity-score matching, we can only conclude correlation and cannot conclusively identify causal directions. Lastly, we acknowledge the limited accuracy on some of our study’s classifiers (see Table \ref{table:miti-empathy-codes}). Although the majority of our therapeutic technique classifiers have high accuracy, the lowest accuracies for MITI classifiers are ~50\% for Seeking Collaboration and ~57\% for Giving Information. Inaccuracies in identifying MITI or empathy codes being used could have reduced our analysis’ predictive accuracy. \textcolor{black}{Although we saw statistically significant results for both classifiers, as discussed in our results, in general we have analyzed any results from these low accuracy MI code classifiers cautiously.}

\section{Conclusion}
We have presented an analysis on the longitudinal and clinical effectiveness of online peer support platforms for depression and anxiety outcomes. Through a pre-post observational study design, we found that having online peer support chats with volunteer counselors improves depression and anxiety symptoms. Additionally, our analysis found that, although online peer support chats are effective, the specific therapeutic behaviors that underlie these platforms, such as support, affirmation, and empathy, do not explain this effectiveness. Our work has important implications for the future of designing online mental health communities and for digital training for online support providers.

\bibliographystyle{ACM-Reference-Format}
\bibliography{ref.bib}

\newpage
\appendix

\section{Full PHQ-2 and GAD-2 questionnaires}

\textbf{PHQ-2}:
Over the last 2 weeks, how often have you been bothered by the following problems?
\begin{enumerate}
    \item Little interest or pleasure in doing things
        \begin{itemize}
   \item Not at all
   \item Several days
   \item More than half the days
   \item Nearly every day
        \end{itemize}
    \item Feeling down, depressed or hopeless
        \begin{itemize}
   \item Not at all
   \item Several days
   \item More than half the days
   \item Nearly every day
        \end{itemize}
\end{enumerate}

\textbf{GAD-2}:
Over the last 2 weeks, how often have you been bothered by the following problems?
\begin{enumerate}
    \item Feeling nervous, anxious or on edge
        \begin{itemize}
   \item Not at all
   \item Several days
   \item More than half the days
   \item Nearly every day
        \end{itemize}
    \item Not being able to stop or control worrying
        \begin{itemize}
   \item Not at all
   \item Several days
   \item More than half the days
   \item Nearly every day
        \end{itemize}
\end{enumerate}

\end{document}